\documentclass{ws-procs9x6}
\usepackage{epsfig}
\usepackage{graphicx}
\input epsf.sty

\def\lsi{\raise0.3ex\hbox{$<$\kern-0.75em\raise-1.1ex\hbox{$\sim$}}}
\def\gsi{\raise0.3ex\hbox{$>$\kern-0.75em\raise-1.1ex\hbox{$\sim$}}}
\newcommand{\lsim}{\mathop{\lsi}}

\begin{document}

\title{Lattice QCD at non-vanishing density: phase diagram, equation of state}

\author{F.~CSIKOR$^{\lowercase{a}}$, G.I.~EGRI$^{\lowercase{a}}$, Z.~FODOR$^{\lowercase{a,b}}$, 
S.D.~KATZ$^{\lowercase{c}}$\thanks{on leave from 
\uppercase{I}nstitute for \uppercase{T}heoretical \uppercase{P}hysics, \uppercase{E}\"otv\"os \uppercase{U}niversity, \uppercase{B}udapest, \uppercase{H}ungary}, K.K.~SZAB\'O$^{\lowercase{a}}$, A.I.~T\'OTH$^{\lowercase{a}}$}

\address{
$^a$Institute for Theoretical Physics, E\"otv\"os University, Budapest, Hungary\\[4mm]
$^b$Department of Physics, University of Wuppertal, Germany\\[4mm]
$^c$Deutsches Elektronen-Synchrotron DESY, Hamburg, Germany}

\maketitle

\abstracts{
We propose a method to study lattice QCD at non-vanishing 
temperature ($T$) and chemical potential ($\mu$). 
We use $n_f$=2+1 dynamical
staggered quarks with semi-realistic masses on $L_t$=4 lattices.
The critical endpoint
(E) of QCD on the Re($\mu$)-T plane is located. 
We calculate the pressure (p), the energy density ($\epsilon$) 
and the baryon density ($n_B$) of QCD at
non-vanishing T and
$\mu$. 
}

\section{Introduction}

QCD at finite $T$ and/or $\mu$ is of fundamental importance,
since it describes  physics relevant
in the early universe, in neutron stars and in heavy ion collisions.
Extensive experimental work has been done
with heavy ion collisions at CERN and Brookhaven to explore
the $\mu$-$T$ phase boundary (cf. \cite{S02}).  Note, that
past, present and future heavy ion
experiments with always higher and higher energies produce states
closer and closer to the $T$ axis of the $\mu$-$T$ diagram. It is
a long-standing question, whether a critical point
exists on the $\mu$-$T$ plane,
and particularly how to tell its location theoretically
\cite{crit_point}.

Universal arguments \cite{PW84} and lattice results \cite{U97}
indicate that at $\mu$=0 
the real world probably has a crossover.
Arguments
based on a variety of models (see e.g. \cite{B89,qcd_phase,crit_point})
predict a first order finite $T$ phase transition at large $\mu$.
Combining the $\mu=0$ and large $\mu$ informations 
suggests
that the phase diagram features a critical endpoint $E$ (with
chemical potential $\mu_E$ and temperature $T_E$), at which
the line of first order phase transitions ($\mu>\mu_E$ and $T<T_E$)
ends \cite{crit_point}. At E the phase transition is of
second order and long wavelength fluctuations appear, which
results in (see e.g. \cite{BPSS01}) consequences, similar to
critical opalescence. Passing close enough to ($\mu_E$,$T_E$)
one expects simultaneous 
signatures
which exhibit non-monotonic dependence on the
control parameters \cite{SRS99},
since one can miss the critical point on either side.

The location of E is
an unambiguous, non-perturbative prediction of QCD.
No {\it ab initio}, lattice QCD study based was done to locate
E.  Crude models
with $m_s=\infty$ were used (e.g. \cite{crit_point})
suggesting that $\mu_E \approx$ 700~MeV, which should be smaller
for finite $m_s$. The goal of our
work is to propose a new method to study lattice QCD at 
finite $\mu$ and apply it to locate the endpoint.
We use full QCD with dynamical $n_f$=2+1 staggered quarks.

QCD at finite $\mu$ can be given
on the lattice \cite{HK83}; however, standard
Monte-Carlo techniques fail. At Re($\mu$)$\neq$0 
the determinant of the Euclidean Dirac operator is complex, which
spoils any importance sampling method.

Several suggestions were studied in detail to solve the problem.

For small gauge coupling an attractive
approach is the ``Glasgow method'' \cite{glasgow} in which the
partition function is expanded in powers of $\exp(\mu/T)$
by using an ensemble of configurations weighted by the $\mu$=0 action.
After collecting more than 20 million configurations only unphysical
results were obtained: a premature onset transition.
The reason is that the $\mu$=0 ensemble does not overlap sufficiently
with the states of interest.
We show how to handle this problem for small $\mu$ values.

At imaginary $\mu$ the measure remains positive and standard Monte Carlo
techniques apply. 
One can also use the fact that the partition function away from the
transition line should be an analytic function of $\mu$, and the fit
for imaginary $\mu$ values could be analytically continued to real
values of $\mu$ \cite{deForcrand:2002ci,D'Elia:2002aa}.  

Due to the renewed interest several promising ideas appeared in the 
last years (without giving a complete list see e.g. 
\cite{Hong:2003fe,Ambjorn:2002pz,Wiese:ws,Liu:2002qr,Akemann:2003by}).

\begin{figure}[htb]
\begin{center}
\includegraphics*[width=6.9cm,]{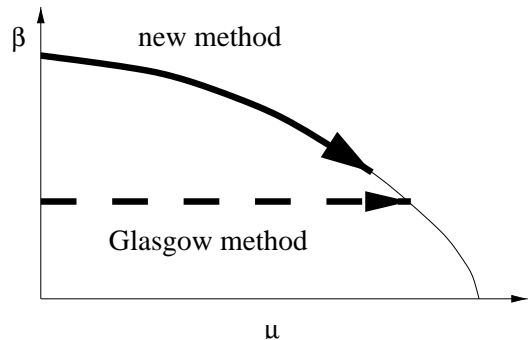}
\end{center}
\caption[]{Schematic difference between the 
present and the Glasgow methods.}
\label{fig1}
\end{figure}

We propose a method
to reduce the overlap problem and determine the
phase diagram in the $\mu$-T plane (for details see 
\cite{Fodor:2001au}).
The idea is to produce an ensemble of QCD configurations at
$\mu$=0 and at the transition temperature $T_c$. Then we determine
the Boltzmann weights \cite{FS89} of these configurations at $\mu\neq 0$
and at $T$ lowered to the transition temperatures at this
non-vanishing $\mu$. Since transition configurations
are reweighted to transition ones a much better
overlap can be observed than by reweighting pure had\-ronic configurations
to transition ones \cite{glasgow}. Since the original 
ensemble is collected at $\mu$=0 we do not expect it to be able to
describe the physics of the large $\mu$ region with e.g. exotic
colour superconductivity. Fortunately, the typical $\mu$ values
at present heavy ion accelerators are smaller than the region
we cover.   

We illustrate the applicability of the method
and locate the critical point of QCD.
(Multi-dimensional reweighting
was successful for determining
the endpoint of the hot electroweak plasma \cite{ewpt}
e.g. on 4D lattices.)
Furthermore we give the equation of state of the QCD plasma
at non-vanishing temperature and chemical potential.

\section{Overlap improving multi-parameter reweighting}

Let us study a generic system of fermions $\psi$ and bosons $\phi$,
where the fermion Lagrange density is ${\bar \psi}M(\phi)\psi$.
Integrating over the Grassmann fields we get:
\begin{equation}\label{path_int}
Z(\alpha)=\int{ D}\phi \exp[-S_{bos}(\alpha,\phi)]\det M(\phi,\alpha),
\end{equation}
where $\alpha$ is the parameter set of
the Lagrangian. In the case of staggered QCD $\alpha$
consists of $\beta$,
$m_q$ and $\mu$.
For some choice of the
parameters $\alpha$=$\alpha_0$
importance sampling can be done (e.g. for Re($\mu$)=0).
Rewriting eq. (\ref{path_int})
\begin{eqnarray}\label{reweight}
Z(\alpha)=
\int { D}\phi \exp[-S_{bos}(\alpha_0,\phi)]\det M(\phi,\alpha_0)&& 
\nonumber \\
\left\{\exp[-S_{bos}(\alpha,\phi)+S_{bos}(\alpha_0,\phi)]
{\det M(\phi,\alpha)  \over \det M(\phi,\alpha_0)}\right\}.&&
\end{eqnarray}
The curly bracket is  
measured on each independent configuration and is interpreted 
as a weight factor $\{ w(\beta,\mu,m,U)\}$. 
The rest is treated as the integration measure. Changing
only one parameter of the ensemble
generated at $\alpha_0$ provides an accurate value for some observables
only for high statistics. This is ensured by
rare fluctuations as the mismatched measure occasionally sampled the
regions where the integrand is large. This is the
overlap problem. Having several parameters
the set $\alpha_0$ can be adjusted to get
a better overlap than obtained by varying only one parameter.

The basic idea of the method as applied to dynamical QCD can be
summarized as follows. We study the system at ${\rm Re}(\mu)$=0 around
its transition point. Using a Glasgow-type technique we calculate the
determinants for each configuration for a set of $\mu$, which, similarly
to the Ferrenberg-Swendsen method \cite{FS89}, can be used for
reweighting.  The average plaquette values can be used to perform an
additional reweighting in $\beta$.  Since transition configurations were
reweighted to transition ones a much better overlap can be
observed than by reweighting pure hadronic configurations to transition
ones as done by the Glasgow-type techniques. 
The differences between the two
methods are shown in Figure \ref{fig1}. 
(Note, that reweighting techniques have very broad applicability. E.g.  
recently it was possible to determine the topological susceptibility
with overlap fermions \cite{Kovacs:2001bx}.) 

\section{The endpoint of $n_f=2+1$ QCD}

In QCD with $n_f$ staggered quarks
one changes the determinants to their $n_f$/4 power in our two
equations. Importance sampling works at some $\beta$ and
at Re($\mu$)=0. Since $\det M$ is complex
an additional problem arises, one should
choose among the possible Riemann-sheets of the fractional power
in eq. (\ref{reweight}). This can be done by using 
\cite{Fodor:2001au}
the fact that at $\mu$=$\mu_w$ the ratio of the determinants is 1 and
it should be a continuous function of $\mu$.

In the
following we keep $\mu$ real and look for the zeros of $Z$
for complex $\beta$.  At a first order phase transition the free
energy $\propto \log Z(\beta)$ is non-analytic.
A phase transition appears only in the V$\rightarrow \infty$ limit,
but not in a finite $V$. Nevertheless, $Z$
has zeros at finite V, 
which are at complex parameters (e.g. $\beta$). For a
system with first order transition these zeros
approach the real axis as V$\rightarrow \infty$ 
by a $1/V$ scaling.
This V$\rightarrow \infty$ limit generates the non-analyticity of
the free energy. For a system with crossover
$Z$ is analytic, and the zeros do
not approach the real axis as V$\rightarrow \infty$.

\begin{figure}[htb]
\begin{center}
\includegraphics*[width=6.9cm,]{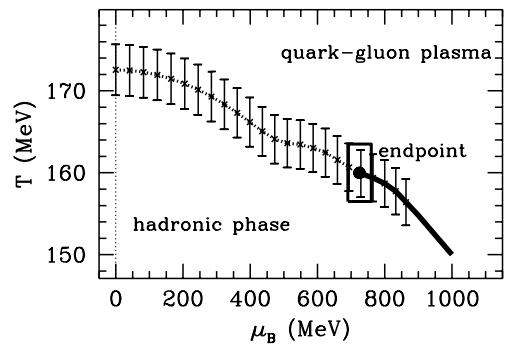}
\end{center}
\caption[]{The T-$\mu$ diagram. Direct results are given with errorbars.
The dotted line shows the crossover, the solid line the first order
transition. The box gives the uncertainties of the endpoint.}
\label{fig4}
\end{figure}

At T$\neq$0 we used $L_t$=4, $L_s$=4,6,8 lattices. T=0 runs were done on
$10^3\cdot$ 16 lattices. $m_{u,d}$=0.025 and $m_s$=0.2 were
our bare quark masses.
At  $T\neq 0$ we determined the complex valued Lee-Yang zeros \cite{LY52},
$\beta_0$, for different V-s as a function of $\mu$. Their
V$\rightarrow \infty$ limit was given by a $\beta_0(V)=\beta_0^\infty+\zeta/V$
extrapolation. We used 14000, 3600 and 840 configurations on
$L_s$=4,6 and $8$ lattices, respectively.
For small
$\mu$ values  Im($\beta_0^\infty$) is inconsistent with
zero, and predicts a crossover.
Increasing $\mu$, the value of Im($\beta_0^\infty$) decreases.
Thus the transition becomes consistent with a first order phase
transition. Our primary result is $\mu_{end}=0.375(20)$ in lattice units.

To set the physical scale we used an
average of $R_0$,  $m_\rho$  and
$\sqrt{\sigma}$.
Including systematics due to
finite V we have
$(R_0\cdot m_\pi)=0.73(6)$, which is at least twice, $m_{u,d}$ is
at least four times
as large as the physical values.

Figure \ref{fig4} shows the phase diagram in
physical units, thus
$T$ as a function of $\mu_B$, the baryonic chemical potential
(which is three times larger than the quark chemical potential).
The endpoint
is at $T_E=160 \pm 3.5$~MeV, $\mu_E=725 \pm 35$~MeV.
At $\mu_B$=0 we obtained $T_c=172 \pm 3$~MeV.

Using a Taylor expansion around $\mu$=0, T$\neq$0 for small $\mu$ 
can be used
to determine the curvature of the phase diagram and to calculate
thermal properties \cite{Allton:2002zi}. 
A 
different method, analytic continuation from imaginary
$\mu$, confirmed also the results of \cite{Fodor:2001au}
on the $\mu$--$T$ 
diagram \cite{deForcrand:2002ci,D'Elia:2002aa}. 

\section{Equation of state at non-vanishing T and $\mu$}

\begin{figure}
\begin{center}
\includegraphics*[width=6.9cm,bb=0 100 413 373]{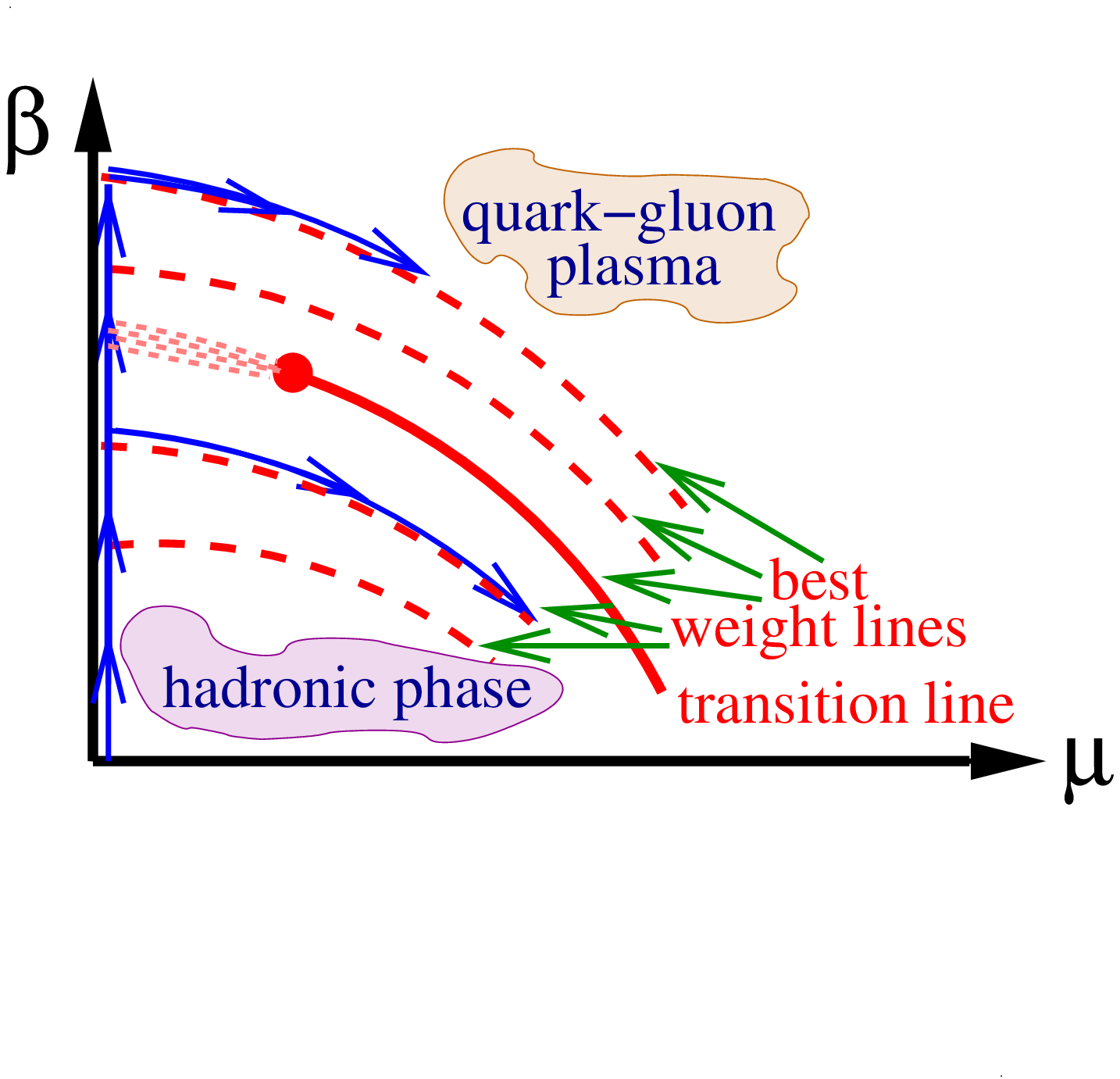}
\end{center}
\caption{\label{weightlines}
The best weight lines on the $\mu$--$\beta$ plane. 
In the middle we 
indicate the transition line. Its first dotted part
is the crossover region. The blob represents the 
critical endpoint, after which the transition is of first order. 
The integration paths used to calculate $p$ are shown by the 
arrows along the $\beta$ axis and the best weight lines.
}
\end{figure}

The equation of state (EOS)
at $\mu$$\neq$0 is essential 
to describe the quark gluon plasma (QGP) formation
at heavy ion collider experiments. Results are only available for $\mu$=0
(e.g. \cite{Gottlieb:1996ae,Karsch:2000ps,AliKhan:2001ek}) at
T$\neq$0.

\begin{figure}
\begin{center}
\includegraphics*[width=6.9cm,bb=0 280 570 700]{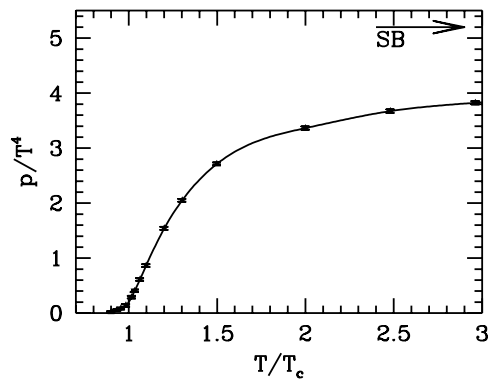}
\end{center}
\caption{\label{eos_p0}
$p$ normalised by $T^4$ as a function of $T/T_c$ at $\mu=0$ 
(to help the continuum interpretation the raw lattice result is multiplied by $c_\mu$=0.446).
}
\end{figure}

\begin{figure}
\begin{center}
\includegraphics*[width=6.9cm,bb=0 280 570 700]{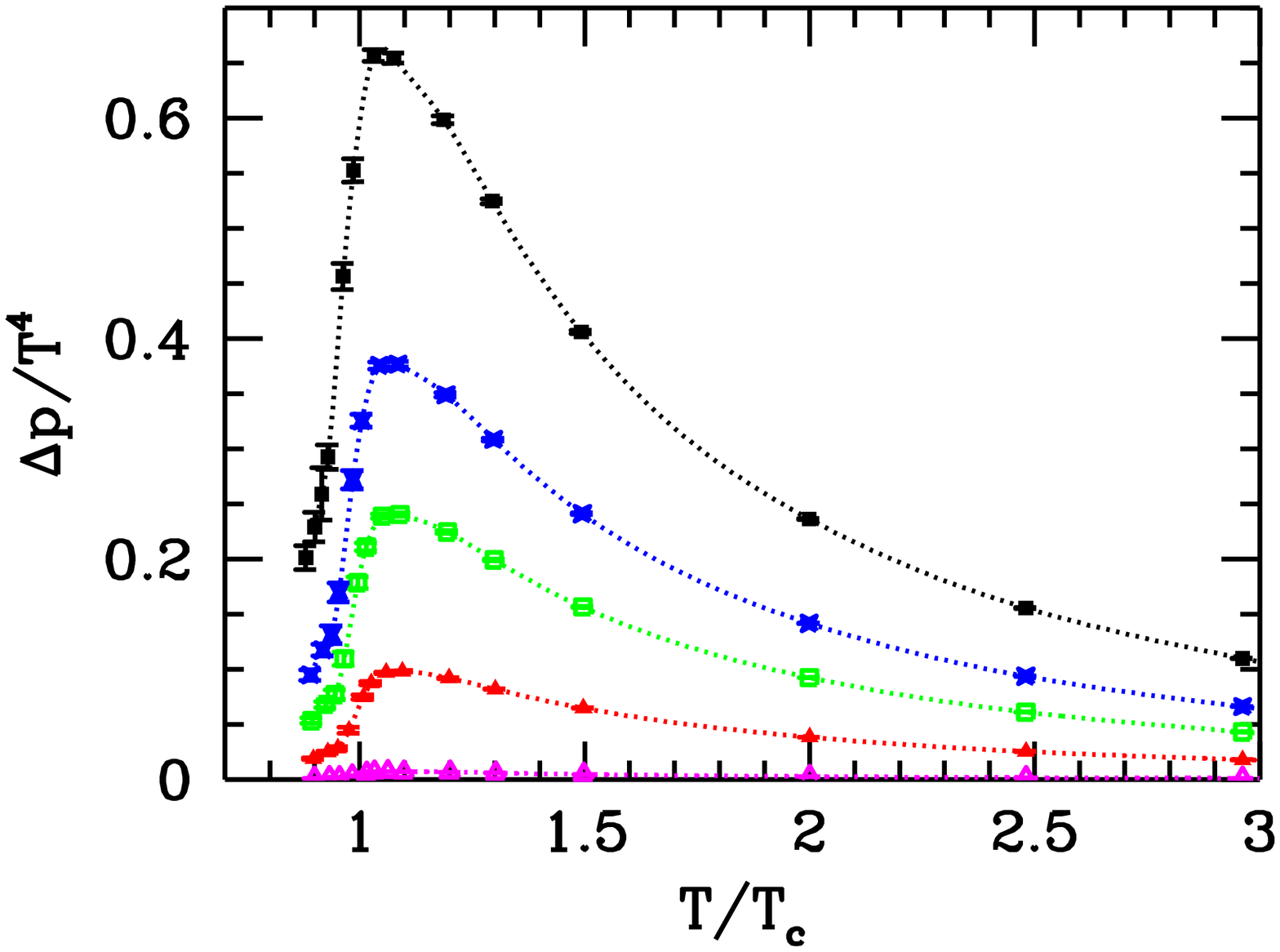}
\end{center}
\caption{\label{eosmu_p_sub}
$\Delta p = p(\mu \neq 0, T)-p(\mu=0, T)$ normalised by $T^4$ as a function of $T/T_c$ for $\mu_B$=100, 210, 330, 410~MeV and 530~MeV (from bottom to top). 
To help the continuum interpretation the raw lattice result is multiplied by $c_{\mu}$=0.446.
}
\end{figure}

We use 4 $\cdot N_s^3$ lattices at $T$$\neq$0 
with $N_s$=8,10,12 for reweighting 
and we extrapolate to V$\rightarrow$$\infty$ 
using the available volumes ($V$). 
At $T$=0 lattices of $24\cdot14^3$ 
are taken for vacuum subtraction and
to connect lattice parameters to physical
quantities. 14 different $\beta$ values are used, which
correspond to $T/T_c=0.8,\dots,3$. Our T=0 simulations 
provided $R_0$ and $\sigma$. The lattice spacing at $T_c$ 
is $\approx$0.25--0.30~fm. We use 2+1 flavours of
dynamical staggered quarks. 
While varying $\beta$ (thus T) we keep the physical
quark masses approx. constant 
(the pion to rho mass ratio is $m_\pi/m_\rho\approx$0.66).

The determination of the equation of state at $\mu\neq$0
needs several observables, ${ O}$,
at $\mu$$\neq$0. This is obtained by using  
the weights of eq. (\ref{reweight})
\begin{equation}
{\overline { O}}(\beta,\mu,m)=\frac{\sum \{w(\beta,\mu,m,U)\} 
{ O}(\beta,\mu,m,U)}{\sum
\{w(\beta,\mu,m,U)\}}.
\end{equation}

$p$ can be obtained from the partition function
as $p$=$T\cdot\partial \log Z/ \partial V$ which can be written as
$p$=(T/V)$\cdot\log Z$ for large homogeneous systems.
On the lattice we can only determine the derivatives of $\log Z$ with respect
to the parameters of the action ($\beta, m, \mu$).
Using the  notation 
$\langle  { O}(\beta,\mu,m) \rangle$= 
${\overline {{ {O}}}(\beta,\mu,m)}_{T\neq0}-
{\overline {{ O}}(\beta,\mu=0,m)}_{T=0}$. 
$p$ can be written as 
an integral \cite{Engels:1990vr}:
\begin{eqnarray}
&&\frac{p}{T^4}=\frac{1}{T^3 V} \int d(\beta, m,\mu ) 
\\
&&\left(
\left\langle \frac{\partial(\log Z)}{\partial \beta}\right\rangle,
\left\langle \frac{\partial(\log Z)}{\partial m}\right\rangle,
\left\langle \frac{\partial(\log Z)}{\partial \mu }\right\rangle\right).
\nonumber
\end{eqnarray}
The integral is by definition independent of the integration path.
The chosen integration paths are shown in Fig \ref{weightlines}. 

The energy density can be written as 
$\epsilon =(T^2/V)\cdot \partial(\log Z)/\partial {T} 
+(\mu T/V)\cdot \partial(\log Z)/\partial\mu$.
By changing the lattice spacing $T$ and $V$ are simultaneously varied.
The special combination $\epsilon-3p$ contains only 
derivatives with respect to $a$ and $\mu$:
\begin{equation}
\frac{\epsilon-3p}{T^4}=-\left.\frac{a}{T^3V}\frac{\partial \log(Z)}{\partial a}\right|_\mu
+\left. \frac{\mu}{T^3 V}\frac{\partial(\log Z)}{\partial\mu}\right|_a.
\end{equation}

\begin{figure}
\begin{center}
\includegraphics*[width=6.9cm,bb=0 280 570 700]{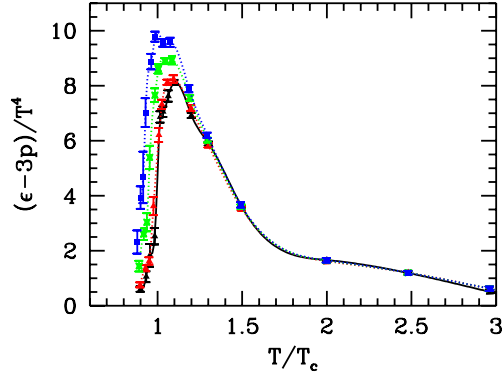}
\end{center}
\caption{\label{interaction}
$(\epsilon-3p)/T^4$ at $\mu_B$=0, 210, 410~MeV and
530~MeV versus $T/T_c$
(from bottom to top). 
To help the continuum interpretation the raw lattice result is multiplied by $c_p$=0.518.
}
\end{figure}

\begin{figure}
\begin{center}
\includegraphics*[width=6.9cm,bb=0 280 570 700]{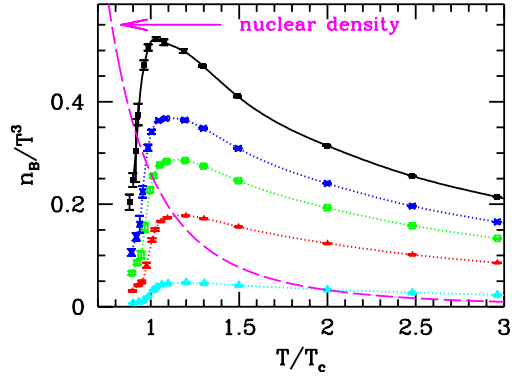}
\end{center}
\caption{\label{density}
$n_B/T^3$ versus $T/T_c$ 
for the same $\mu_B$ values as in Fig. 3 
(from bottom to top).
(to help the continuum interpretation the raw lattice result is multiplied by
$c_\mu$=0.446). 
As a reference value the line starting in the left upper corner 
indicates the nuclear density. 
}
\end{figure}

The quark number density is $n=(T/V)\cdot \partial \log(Z)/\partial \mu$ which 
can be measured directly or obtained from $p$ 
(baryon density is $n_B$=$n$/3 and baryonic chemical 
potential is $\mu_B$=3$\mu$).

We present direct lattice results on $p(\mu=0,T)$, 
$\Delta p(\mu,T)=p(\mu\neq 0,T)-p(\mu=0,T)$, $\epsilon(\mu,T)$-3$p(\mu,T)$ and 
$n_B(\mu,T)$. Additional overall factors
were used to help the phenomenological interpretation.
 
Fig. \ref{eos_p0} shows $p$  at $\mu$=0. 
In Fig. \ref{eosmu_p_sub} we show $\Delta p/T^4$ for different 
$\mu$ values. 
Fig. \ref{interaction} shows $\epsilon$-3$p$ normalised by
$T^4$, which tends to zero for large $T$. 
Fig. \ref{density} gives the baryonic density
as a function of $T/T_c$ for different $\mu$-s.  
The densities can exceed the nuclear density
by up to an order of magnitude.

An important finding concerns  
the applicability of our reweighting
method: the maximal $\mu$  
scales with the volume as 
$\mu_{\rm{max}}\cdot a \sim (N_t\cdot N_s^3)^{-0.25}$.
If this behaviour persists, one could --in principle--
approach the true continuum limit   
($a \sim 1/N_t \sim (N_t\cdot N_s^3)^{-0.25}$, thus
$\mu_{\rm{max}}$$\approx$const.).

\section{Conclusion}

We proposed a method --an overlap improving multi-parameter reweighting
technique-- to numerically study non-zero $\mu$ and determine the
phase diagram in the $T$-$\mu$ plane.
Our method is applicable to any number of Wilson or staggered quarks.

We studied the $\mu$-$T$ phase diagram of QCD with
dynamical $n_f$=2+1 quarks.
Using our method we obtained
$T_E$$\approx$160~MeV and $\mu_E$$\approx$700~MeV for the endpoint.
Though $\mu_E$ is too
large to be studied at RHIC or LHC, the endpoint would
probably move closer to the $\mu$=0 axis
when the quark masses get reduced.

The equation of state was determined on the temperature versus
chemical potential plane. According to our results the applicability
range of the overlap improving multi-parameter reweighting method
for the quark chemical potential can be summarized as $\mu\lsim T$.

Clearly, more work is needed to get
the final values by extrapolating
in the R-algorithm and to the thermodynamic, chiral and continuum limits.
The details of the presented results can be found in 
\cite{Fodor:2001au,Csikor:2002aa}.

\section*{Acknowledgements}

This work was partially supported by Hungarian Scientific
grants OTKA-T37615/\-T34980/\-T29803/\-M37071/\-OMFB1548/\-OMMU-708. 
For the simulations a modified version of the MILC
public code was used (see http://physics.indiana.edu/\~{ }sg/milc.html). 
The simulations were carried out on the 
E\"otv\"os Univ., Inst. Theor. Phys. 163 node parallel PC cluster.

\end{document}